\documentclass[apjl]{emulateapj}

\usepackage{graphicx}
\usepackage{enumerate}
\usepackage[english]{babel}
\usepackage[T1]{fontenc}
\usepackage{hyperref}
\usepackage{amsmath}
\usepackage{apjfonts}
\usepackage{natbib}
\newcommand{\myemail}{matteo.cerruti@cfa.harvard.edu}
\newcommand{\dermeremail}{charles.dermer@nrl.navy.mil}
\newcommand{\threec}{\textit{3C~454.3}}
\newcommand{\threecbis}{\textit{3C~279}}
\newcommand{\fermi}{\textit{Fermi}}
\newcommand{\fermilat}{\textit{Fermi-LAT}}

\begin{document}

\title{Gamma-ray blazars near equipartition and the  origin of the GeV spectral break in \threec}
\shorttitle{Origin of GeV spectral break in \threec}

\author{Matteo Cerruti\altaffilmark{1,2}, Charles D. Dermer\altaffilmark{3}, Beno\^{i}t Lott\altaffilmark{4}, Catherine Boisson\altaffilmark{2} and  Andreas Zech\altaffilmark{2} }

\altaffiltext{1}{Harvard-Smithsonian Center for Astrophysics; 60 Garden Street, 02138 Cambridge, MA, USA.\\  \myemail    }
\altaffiltext{2}{LUTH, Observatoire de Paris, CNRS, Universit\'{e} Paris Diderot; 5 Place Janssen, 92190, Meudon, France}
\altaffiltext{3}{Code 7653, Space Science Division, U.S. Naval Research Laboratory; 20375, Washington, DC, USA.\\ \dermeremail}
\altaffiltext{4}{CENBG, Universit\'{e} de Bordeaux, CNRS/IN2P3, UMR 7595, Gradignan, 33175, France}

\begin{abstract}
Observations performed with the \fermilat\ telescope have revealed the presence of a spectral break in the GeV spectrum of flat-spectrum radio quasars (FSRQs) and other low- and intermediate-synchrotron peaked blazars. We propose that this feature can be explained by Compton scattering of broad-line region (BLR) photons by a non-thermal population of electrons described by a log-parabolic function. We consider in particular a scenario in which the energy densities of particles, magnetic field, and soft photons in the emitting region are close to equipartition. We show that this model can satisfactorily account for the overall spectral energy distribution of the FSRQ \threec, reproducing the GeV spectal cutoff due to Klein-Nishina effects and a curving electron distribution.\\
\end{abstract}


\section{Introduction}

The Large Area Telescope (\textit{LAT}) on board the \fermi\ Gamma ray Space Telescope \citep{FermiLAT} has significantly improved our knowledge of the properties of active galactic nuclei (AGN) emitting GeV photons. The science of blazars, which represent more than 95\% of sources in the \fermilat\ extragalactic catalog \citep{Fermiagncatalog}, has particularly benefited thanks to \fermi\ and its all-sky observation mode. The main observational characteristics of blazars are extreme variability, a high degree of polarization, and a spectral energy distribution (SED) dominated by a non-thermal continuum at all wavelengths \citep[see e.g.][]{Urry95}. 
Understanding blazars therefore depends on simultaneous multiwavelength campaigns. Since the launch of the \fermi\ satellite, the $\gamma$-ray community has, for the first time, access to uninterrupted light-curves and spectral measurements of dozens of blazars in the GeV energy band.

Blazar SEDs generally exhibit two broad components. The lower-energy one is commonly ascribed to non-thermal synchrotron radiation, and peaks in a $\nu F_\nu$ representation between {mm} and X-rays, while the second one, associated in leptonic models with an inverse Compton process, peaks at $\gamma$-ray energies \citep[see e.g.][]{FermiSED}. BL Lac objects show a variety of synchrotron-peak frequencies $\nu_{pk}$, and we can thus differentiate between low-synchrotron-peaked sources \citep[LSP, showing a peak frequency below $10^{14}$ Hz; see][]{fermiagn}, intermediate-synchrotron-peaked sources (ISP,  $10^{14}\leq \nu_{pk} < 10^{15}$ Hz), and high-synchrotron-peaked (HSP) sources, with $\nu_{pk} > 10^{15}$ Hz. In contrast to BL Lac objects, FSRQs are essentially all LSP blazars.

The blazar emission arises from a relativistic jet of plasma with Doppler factor $\delta$ and magnetic field $B$ that contains a non-thermal population of electrons and positrons. The synchrotron emission is thought to make the low-energy component of the SED, while low-energy photons Compton-scattered to $\gamma$-ray energies make the high-energy component. The soft photons can be the synchrotron emission itself \citep[synchrotron-self-Compton model, SSC,][]{Konigl81} or an external photon field (external-inverse-Compton model, EIC), such as the emission from the BLR \citep{Sikora94}, the dust torus \citep{Blaz00} or the accretion disk \citep[][]{Dermer93}. The SSC model can satisfactorily describe the SED of HSP BL Lac objects, while for LSP blazars an external photon field is required \citep[e.g.,][]{Ghisellini11}. 

The modeling of blazars is, however, complicated by the large number of free parameters, especially for the EIC scenario. In a companion paper \citep{equipartitionpaper}, we have simplified the modeling by introducing equipartition relations between the energy densities of particles, photons and magnetic field, and by parameterizing the particle distribution with a log-parabolic function. While in that paper we focus on the properties of the overall SEDs produced in the equipartition scenario, and on the implications from modeling the blazar \threecbis, here we concentrate on the spectral break observed between $\approx1$ and $5$ GeV in the \fermilat\ spectrum of \threec\ and, indeed, all LSP and ISP blazars with sufficiently good statistics  \citep{LBAS,fermibsl,FermiSED}.

In this Letter we show that in the near-equipartition approach, a softening in the GeV spectrum is naturally produced from Klein-Nishina effects when jet electrons scatter BLR radiation, and is consistent with observations of the GeV break.

\section{\threec\ and GeV Spectral Breaks }
\label{section3c}

The $\gamma$-ray emission of \threec\ (2251+158) at redshift $z = 0.859$, was first observed with \textit{EGRET} \citep{Hartman93}. Major pre-\fermi\ outbursts occurred in 2005 \citep{Villata06, Pian06, Fuhrmann06}, and  a large flare in 2007 was observed in $\gamma$-rays with the AGILE experiment \citep{Ghisellini07, Vercellone08, Raiteri08, Vercellone09, Donnarumma09}. Since the launch of the \fermi\ mission, \threec\ has been monitored in the GeV energy band, providing the first continuous long-term light-curve in $\gamma$-rays. The source exhibited several major flares in December 2009 and April 2010 \citep{Ackermann10}, and in November 2010 \citep{Abdo11, Vercellone11}. 

The \fermilat\ observations have shown that the spectrum of this source does not follow a power-law behavior, but instead shows a clear break at an energy of a few GeV \citep{Ackermann10,Abdo11}. Especially interesting is that the energy of the break changes by less than a factor of two when the $>100$ MeV flux changes by nearly a factor of $20$. While the presence of a break is statistically significant, it is  difficult to differentiate between a broken power-law model or a curved function, such as a power-law with an exponential cut-off or a $\gamma$-ray spectrum described by a log-parabolic function, over relatively short periods of time \citep{Abdo11}. Data collected over longer month-long periods yield better statistics, but are in turn insensitive to short-term spectral variations. 

\citet{Harris12} showed that the mean spectrum of \threec\ is best-fitted by a broken power-law function, even though the value of the break energy varies according to the energy range in which the fit is performed, suggesting ``the presence of a curvature in the spectrum [...] as well as any break." On the other hand, \citet{Abdo11} showed a complex spectral behavior for the November 2010 flare, with no function clearly being preferred over the others.

\section{Description of the code}
\label{section2}
The modeling of the SED of \threec\ has been performed using a one-zone SSC model as described in \citet{Katarzynski01}, adding the computation of the EIC emission component for low-energy target photons from the BLR, the dust torus, and the accretion disk. The particle energy
distribution in the emitting region is given by a log-parabolic function. The jet parameters $\delta$, $B$, $R$, and the energy densities of the external radiation field are deduced from the equipartition relations given in \citet{equipartitionpaper}.

\subsection{EIC emission}
\label{threecmodel}

We computed the EIC emission following the formulae described in \citet{Dermerbook}. With respect to the work presented in \citet{equipartitionpaper}, in which we considered only the EIC emission using as a target photon field the Ly$\alpha$ line and the dust torus photons ({with $E\cong0.1$ eV}), here we consider a more complex spectrum of emission lines. The strength of the lines, expressed as a ratio of line fluxes compared to the flux of the dominant Ly$\alpha$ emission line, has been fixed using the ratio-estimation provided by \citet{Telfer02} (see Table \ref{tablelines}). We tested other line-ratio combinations \citep{Francis91}, which provide similar results and do not significantly modify the modeling.

\begin{table}[t!]
\begin{center}
\caption{BLR emission lines included in the modeling of \threec}
\label{tablelines}
\begin{tabular}{c|c c}
\hline
\hline
Line & Flux & E (eV) \\
\hline
Ly$\alpha$ & 100 & 10.20\\
C IV & 52.0 & 8.00 \\
$\textit{Broad feature}$ & 30.2 & 7.75 \\
Mg II & 22.3 & 4.43 \\
N V & 22.0 & 10.00\\
O VI + Ly$\beta$ & 19.1 & 12.04\\
C III + Si III & 13.2 & 6.53\\
\hline
\hline
\end{tabular}
\tablecomments{Line strengths are expressed as a ratio of the line flux to the Ly$\alpha$ flux, as given by \citet{Telfer02}.}
\end{center}
\end{table}

We also verified the accuracy of approximating the external photon field as monochromatic for both the Ly$\alpha$ and the dust torus photons, as assumed in \citet{equipartitionpaper}. Using Gaussian line profiles for the emission lines or  a blackbody function for the torus emission, as used in the following calculations, does not modify significantly the results, which remain valid in the monochromatic approximation used in the companion paper. We furthermore modeled the EIC emission from the accretion-disk photons. The importance of this component depends on the distance of the blazar emitting-region from the disk. In the present treatment, we focus on a scenario where the direct accretion disk photons can be neglected, which constrains the minimum distance of the $\gamma$-ray emitting region from the central nucleus.

\begin{table*}[hbtp]
\begin{center}
\caption{Parameters used for the modelling of \threec.}
\label{tableone}
\begin{tabular}{c|c c c c c c c c| c c c c c c c }
\hline
\hline
 & \multicolumn{8}{c|}{Input} & \multicolumn{7}{c}{Output} \\
\hline
Epoch$^a$ & $L_{48}$ & $t_4$ & $\nu_{14}$ & $\zeta_e$ & $\zeta_s$ & $\zeta_{Ly\alpha}$ & $\zeta_{IR}$ & $b$ & $\delta$ & $B$ & $R$ & $\gamma'_p$ & $N'_e(\gamma'_p)$ & $u_{Ly\alpha}$ & $L_{jet}^b$ \\
 &  & &  &  & & & & & & G & $10^{16}\ \textrm{cm}$ & & $\textrm{cm}^{-3}$ & $10^{-4}\ \textrm{erg cm}^{-3}$ & $10^{45}\ \textrm{erg s}^{-1}$\\
\hline
A & 0.7 & 10  & 0.03 & 0.6 & 0.07 & 1.3  & 1.04 & 1.0 & 22.3 & 0.76 & 6.69& 205 & 0.15 & 1.82& 8.6\\
B & 2.4 & 3.5 & 0.03 & 3.5 & 0.12 & 10.5 & 8.4  & 1.0 & 39.3 & 0.56 & 4.13& 180 & 0.56 & 2.56& 29.1\\
\hline
\hline
\end{tabular}
\endcenter
$^a$Epochs A and B represent the low and high states of 2008 and 2010, respectively. Data and model SEDs are shown in Fig. \ref{fig1}.\\
$^b$Total jet power assuming the energy density of hadrons equals that of electrons.
\end{center}
\end{table*}

\begin{figure*}[hbtp]
\vspace{0.3cm}
\includegraphics[width=265pt]{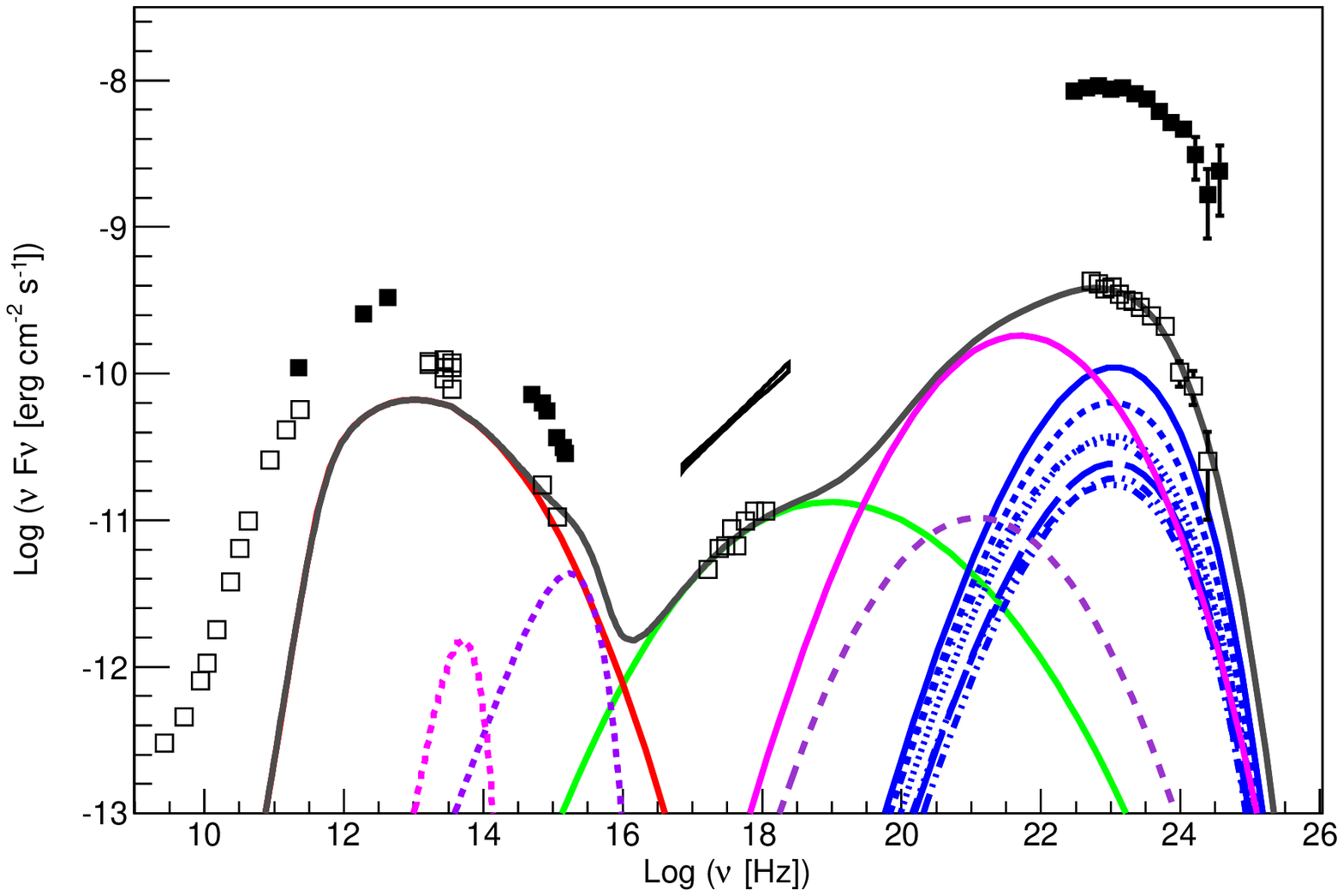}
\setlength{\unitlength}{0.01\linewidth}
    \begin{picture}(0,0)    % picture environment for inset
       \put(-44,23){\includegraphics[width=25\unitlength]{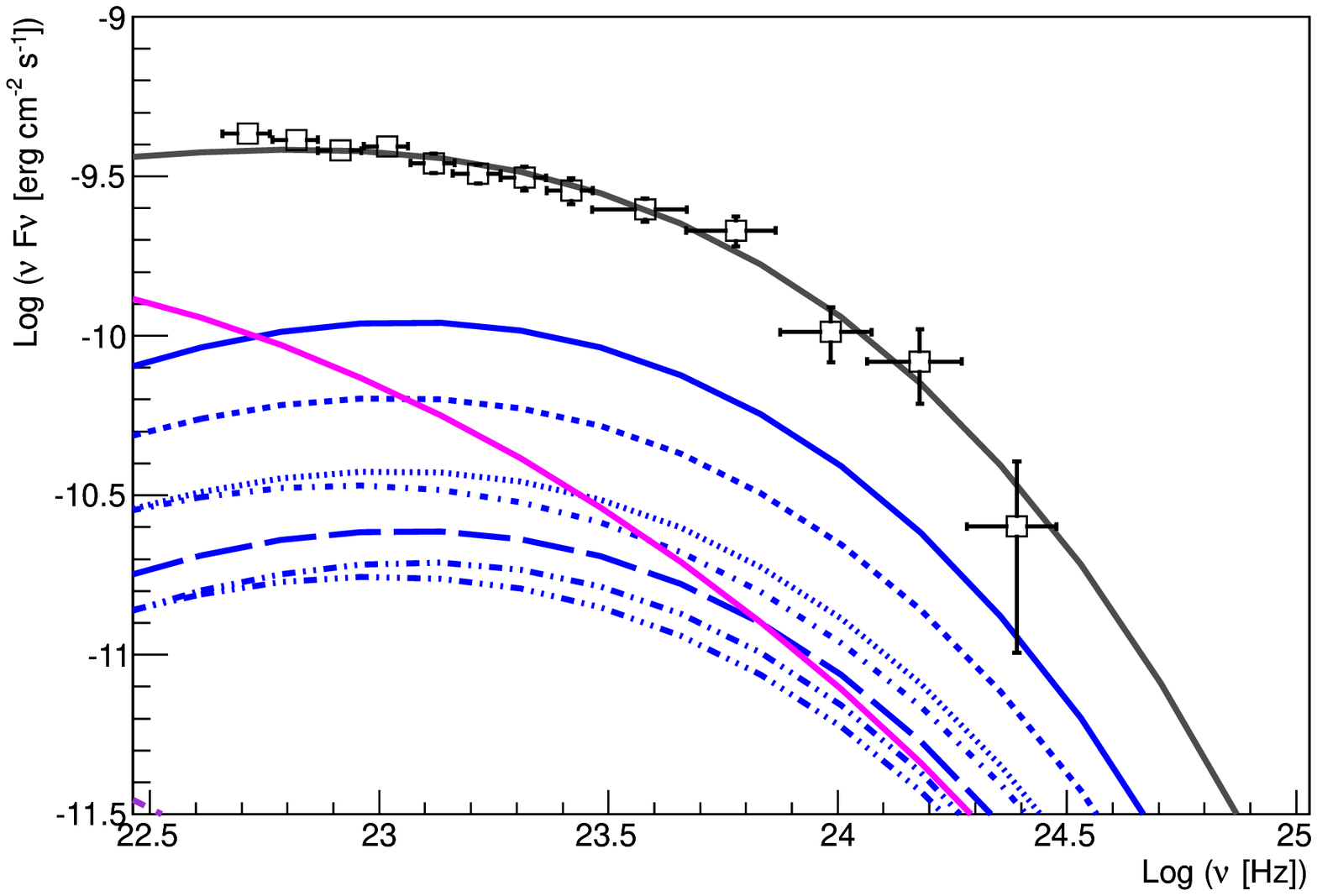}}
    \end{picture}
\includegraphics[width=265pt]{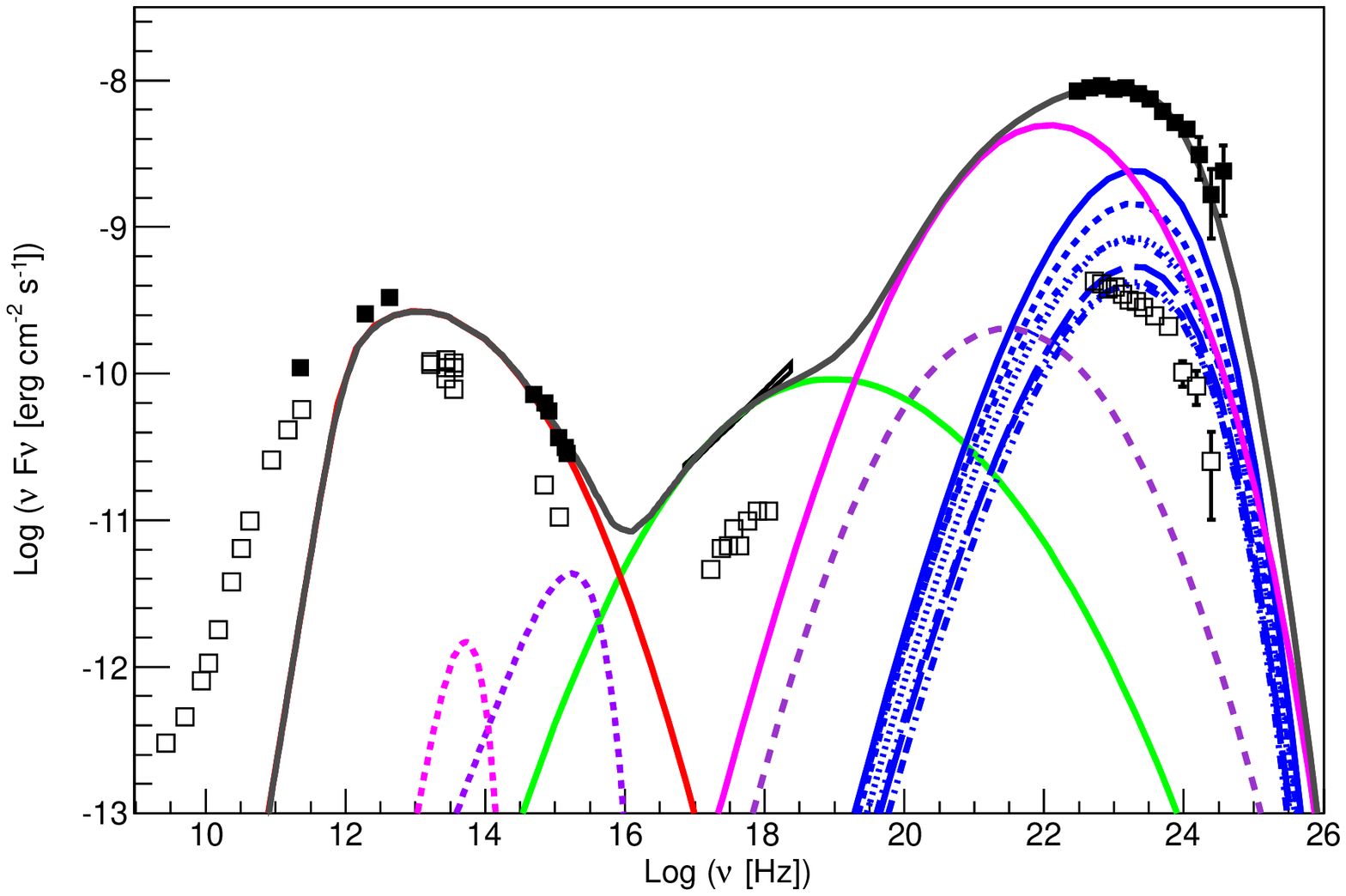}
   \begin{picture}(10,0)    % picture environment for inset
        \put(59.4,24){\includegraphics[width=25\unitlength]{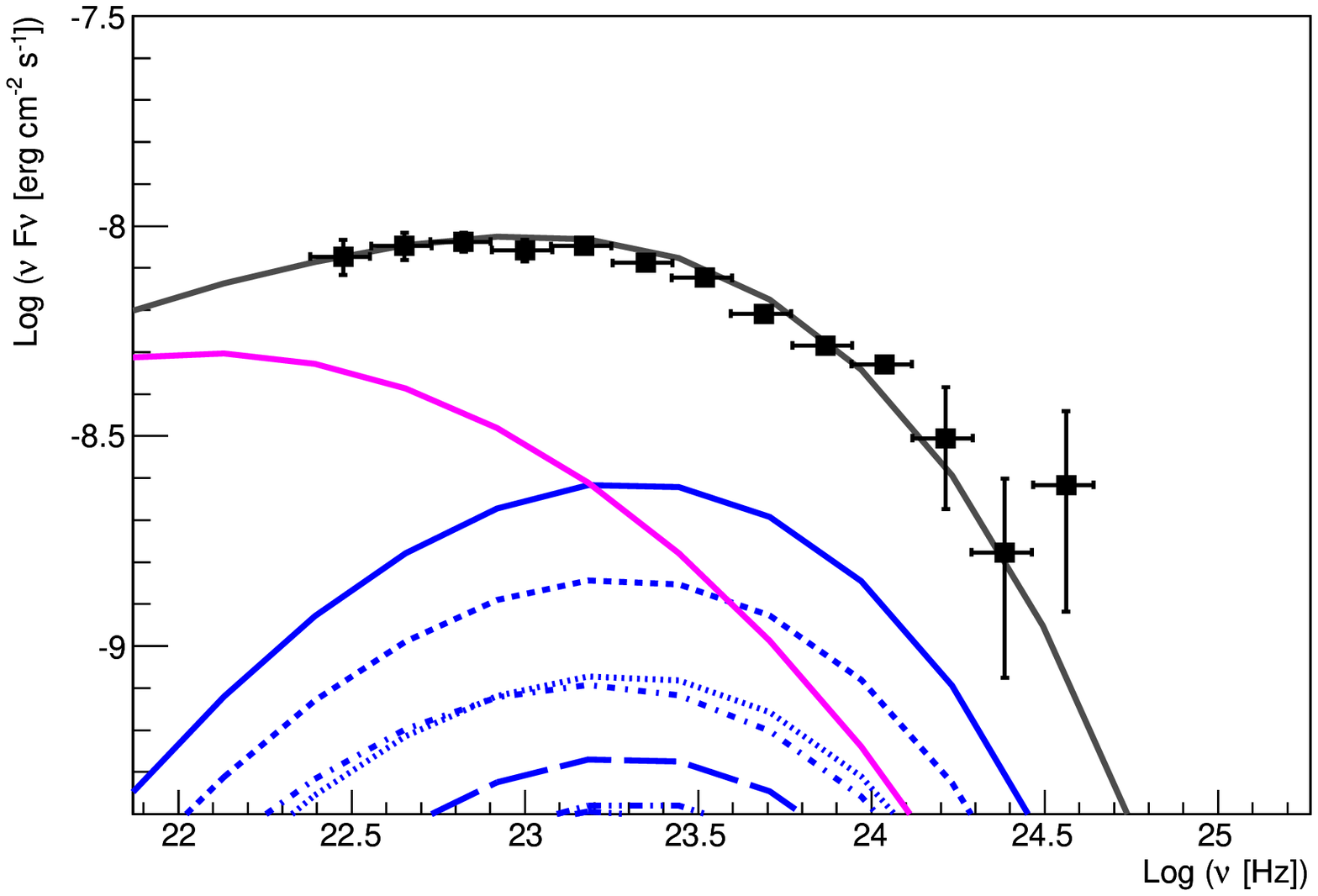}}
    \end{picture}
\caption{Spectral energy distribution of \threec\ observed in August 2008 \citep[open squares, see][]{Abdo09} and November 2010 \citep[full black squares, see][]{Wehrle12}. The model components are, from low to high frequencies, the synchrotron emission (red thick line), the dust torus (magenta dotted line) and accretion disk (violet dotted line) thermal emission, the self-Compton emission (green thick line), the inverse Compton emission over the dust torus photons (magenta thick line), the inverse Compton emission over the accretion disk (violet dashed line) and the inverse Compton emission over a spectrum of emission lines, using the ratio values of \citet{Telfer02} (blue lines, the highest one being the Ly$\alpha$). The inset shows a zoom over the \fermilat\ spectrum. The accretion disk and dust torus thermal emissions have been computed assuming  $L_{disk}=3.6\cdot 10^{46}$ erg s$^{-1}$ (see Section \ref{threecres}), and $L_{IR}=0.2\ L_{disk}$. \textit{Left:} modeling of the lower flux state, assuming the parameters provided in Table \ref{tableone}, epoch A. \textit{Right:} modeling of the higher flux state, assuming the parameters provided in Table \ref{tableone}, epoch B. \label{fig1}  \\ \vspace{0.2cm}}
\end{figure*}

\subsection{Log-parabola function and equipartition relations} 

We consider a scenario where the particle distribution is described by a log-parabolic function, 
with model parameters for the energy densities of the photons, magnetic fields, and particles
defined through equipartition relations. The differential electron number density $N^\prime_e(\gamma ')$ is defined through the relation
\begin{equation}
\gamma'^2 N'_e(\gamma')=\gamma'^2_p N'_e(\gamma'_p)\left( \frac{\gamma'}{\gamma'_p} \right)^{-b\ \log{(\gamma'/\gamma'_p)}}
\end{equation}
where $\gamma'$ is the particle Lorentz factor, $b$ is the curvature index, $\gamma'_p$ is the peak Lorentz factor in the $\gamma'^2 N'_e(\gamma')$ distribution, and the primes underline the fact that all quantities are expressed in the blob comoving frame. The nonthermal electron distribution is thus characterized by three free parameters, namely $b$, $\gamma'_p$ and $N'_e(\gamma'_p)$.

The input parameters used to describe the blazar SED are  (i) $L_{48}= L_{syn}/10^{48}~ \textrm{erg s}^{-1}$, giving the apparent luminosity of the synchrotron component;
(ii) $t_4=t_v^{obs}/[(1+z)10^4\ \textrm{s})]$, giving the source variability timescale in terms of the measured variability timescale $t_v^{obs}$;
(iii) $\nu_{14}=(1+z){\nu_{syn}^{obs}}/{10^{14}\ \textrm{Hz}}$, giving the peak synchrotron frequency in the source frame in terms of the measured peak synchrotron frequency $\nu_{syn}^{obs}$;
(iv) $\zeta_e={u'_e}/{u'_B}$, the equipartition factor relating the nonthermal electron $(u'_e)$ and magnetic-field $(u'_B)$ energy densities;
(v) $\zeta_s= {u'_s}/{u'_B}$, the equipartition factor relating the synchrotron $(u'_s)$ and magnetic-field energy densities;
(vi)  $\zeta_{Ly\alpha}={u'_{Ly\alpha}}/{u'_B}$, the equipartition factor relating the Ly$\alpha$ ($u'_{Ly\alpha}$) and magnetic-field energy densities; and
(vii) $b$, the curvature index of the particle distribution, noted above. The presence of hadrons does not affect the spectral model, but will increase the jet power. 
In our calculations, we assume that the energy density of protons and ions is equal to the energy density of electrons.

These input parameters are used to deduce model parameters, namely the Doppler factor $\delta$, the comoving fluid magnetic field $B$, the comoving emitting region size $R=c\delta t_v^{obs}/(1+z)$, and the principal electron Lorentz factor $\gamma'_p$, 
using the equations derived in \citet{equipartitionpaper}. The normalization of the particle distribution and of the external photon field are obtained through $\zeta_e$ and $\zeta_*$, while $b$ is directly used as input for the code. In our calculation, we assume that the Doppler factor is related to the bulk Lorentz factor through $\delta=2\Gamma$. This assumption is valid only if the jet is seen on-axis ($\theta_{obs}\simeq 0$). For the case of \threec, \citet{Jorstad05} estimated $\theta_{obs}$ comprised between $0.2^\circ$ and $3.9^\circ$, for different radio components. The component with the lowest value of $\theta_{obs}$ is remarkably stationary over several years with respect to the position of the radio-core. Assuming instead a value of $\delta=\Gamma$ (corresponding to $\theta_{obs}\simeq 1/\Gamma$) does not affect the synchrotron and SSC emissions, while the normalization of the EIC components is reduced. The higher value of $\Gamma$ implies as well a higher value of the jet power $L_{jet}\approx 2\pi R^2 \beta \Gamma^2c u'_{tot}$, where $u'_{tot}$ is the sum of all the energy densities \citep[see, e.g., ][]{Celotti08}. However, this particular choice does not affect the modeling of the spectral break in the GeV energy range, which is the main purpose of this Letter.

\section{SED Modeling}
\label{threecres}

The SED of \threec\ is shown in Fig. \ref{fig1}. Given the extreme variability observed in the $\gamma$-ray emission from this source, we have focused on two different states: a relatively low-state taken from \citet{Abdo09} (observations performed in August 2008, hereafter epoch A), and a high state taken from \citet{Wehrle12} (corresponding to the highest flux state, observed in November 2010, epoch B).\footnote{For the high-flux state, the \fermilat\ data obtained during MJD 55520 have been analyzed independently from \citet{Wehrle12}. The \fermilat\ data points shown in Fig. \ref{fig1} come from our own analysis, giving results identical to that performed in \citet{Abdo11}.}  The UV data are modeled by the declining, high-energy part of the synchrotron component, which  peak at infrared frequencies. The blue bump emissions from the accretion disk and BLR can contribute to the UV flux measured in epoch A, while in epoch B the UV spectrum is dominated by the nonthermal continuum. The X-ray data represent, on the other hand, the rising part of the IC component, and in this scenario its origin is ascribed to the SSC component. The different EIC components produce the $\gamma$-rays observed by \fermilat. Radio data are considered in both cases as being made by extended regions beyond the inner jet considered here, which is in accord with the lack of radiation produced at these frequencies as a consequence of synchrotron self-absorption. 

The curvature parameter $b$ of the electron distribution fits both SEDs with a value of unity. The frequency of the synchrotron peak in both epochs is assumed to be $\nu_{14} = 0.03$, though we could adjust that value if the data quality in the mm-IR regime required a different value.  The variability timescale for epoch A is taken to be $t_4=10$, while the variability timescale for epoch B is constrained by the $\gamma$-ray variability of the flaring state and set equal to $t_4=3.5$, which corresponds to an observed variability timescale of roughly $18$ hours \citep[see the \fermilat\ lightcurve in][ Fig.6]{Wehrle12}. The other input parameters differ between the two epochs and are reported in Table \ref{tableone}. In particular, the synchrotron luminosity  varies from $L_{48} = 0.7$ to $L_{48} = 2.4$ between epochs A and B. The ratio of the energy density of the dust to the Ly$\alpha$ radiation has been fixed in both cases to $\zeta_{IR} / \zeta_{Ly\alpha}=0.8$, constrained by the low-energy \fermi\ data, and consistent with the results presented by \citet{equipartitionpaper} for the case of \threecbis.

As can be seen in Fig. \ref{fig1}, this scenario reproduces the \fermilat\ observations for both the lower and the higher flux states. Note that even though the electron energy distribution is parameterized by a log-parabola function, the emission in the GeV band is not necessarily log-parabolic, given in particular the transition to the Klein-Nishina regime, and the superposition of contributions from different external photon fields (dust and lines from the BLR) with different temperatures.

Looking at the output parameters, we see that the emitting region in the high state is characterized by a larger Doppler factor ($39$ vs.\ $22$) and a smaller radius ($4\times 10^{16}$ cm vs.\ $7\times 10^{16}$ cm) and magnetic field ($0.6$ G vs.\ $0.8$ G) with respect to the lower state. In order to describe the \fermi\ spectrum, the energy density of the external BLR photon field increases by roughly $40\%$ from epoch A to epoch B. 

Our modeling can be compared with the one done by \citet{Bonnoli11}. In their model, the EIC component is more important than the SSC, and contributes significantly at soft X-ray energies as well. They obtain higher magnetic field values (of the order of $4$-$6$ G), and Doppler factors close to the value we derive for epoch A, and never higher than $\delta = 30$. The emitting region size they find is slightly smaller ($10^{16}$ cm), but this parameter is strongly dependent on the assumed variability timescale.

The EIC scattering of the thermal photons emitted by the accretion disk depends on the assumptions made on the mass of the supermassive black hole $M_\bullet$, the Eddington luminosity ratio $l_{Edd}$, the accretion efficiency $\eta$, and the location of the emitting region $r_\gamma$. In the plots shown in Fig.\ \ref{fig1} we have assumed $M_\bullet=5\cdot 10^{8}\ M_\odot$ \citep{Bonnoli11}, $l_{Edd}=0.5$, $\eta=0.1$, and $r_\gamma=10^4\ R_G$. In particular, if the emitting region is located closer to the accretion disk, this component becomes more important, and it starts dominating the SED when $r_\gamma \lesssim 3000\ R_G$. 

The location of the $\gamma$-ray emitting region can be constrained as well by looking at the energy densities of BLR photons \citep[see discussion in][]{equipartitionpaper}. In fact, the radius of the BLR ($R_{BLR}$) can be inferred from the luminosity of the accretion disk, and then used to estimate the BLR photon energy density \citep[see e.g.][]{Ghisellini08}. For the case of \threec, following \citet{Bonnoli11} we assumed $R_{BLR}=6\cdot 10^{17}$ cm ($\simeq 8\cdot 10^3 R_G$) and $L_{disk}=3.6\cdot 10^{46}$ erg s$^{-1}$, which yields $u_{BLR}\simeq 0.03$ erg cm$^{-3}$. The values we derive in our modeling are two orders of magnitude lower than this estimated value, implying an emitting region located farther away, at the outer edge of the BLR, consistent with the locations deduced for the emission regions in \threecbis. 
 
A point worth noting is the prediction of our model for the spectral behavior in the energy range between X-rays and \fermi\, where observations are lacking. In this band, EIC radiation associated with Compton-scattered torus photons dominates the flux, already indicated by Swift BAT observations \citep{Ajello12}. The energy density of the IR radiation field, fixed in our models to $0.8$ times the Ly$\alpha$ one, can be more tightly constrained by better hard-X-ray data.  \textit{NuSTAR} \citep{NuSTAR} is now providing data that can help reveal this missing piece of the SED, as will \textit{Astro-H} \citep{AstroH}.

\section{Discussion and Summary}

The GeV spectral break in LSP and ISP blazars was not predicted, and remains one of the most interesting features in the blazar $\gamma$-ray spectrum. Various theoretical models have been developed to explain the GeV spectral break. After the original discovery of the GeV break in \threec\, \citet{Abdo09} speculated that it was due to a break in the electron spectrum, but in this case the break energy should be strongly dependent on $\delta$ and the source flux.  \citet{Finke10} proposed a model in which the \fermilat\ spectrum is ascribed to the superposition of an EIC emission component from target accretion-disk photons making most of the softer $\gamma$-rays in the GeV spectrum, and EIC radiation from target Ly$\alpha$ photons dominating the emission at energies above the break.  \citet{Poutanen10} proposed an intriguing scenario, developed further in \citet{Stern11}, where the GeV break originates from $\gamma$-$\gamma$ pair-production attenuation when $\gamma$ rays interact with high-ionization photons, including He Ly$\alpha$ photons. This requires the production of a power-law $\gamma$-ray spectrum made deep within the BLR. \citet{Harris12} have shown, however, that the break energies are not consistent with the ones predicted by \citet{Stern11}, weakening such an absorption scenario.

We have considered in detail a scenario originally suggested by \citet{Ackermann10}, where it was noticed that Klein-Nishina effects on Compton scattering of target 10.2 eV Ly$\alpha$ photons, the dominant BLR emission line, naturally makes a break at a few GeV, independent of $\delta$. As shown there, a simple power-law electron distribution scattering the Ly$\alpha$ photons results in a spectrum too hard with respect to the spectral data.  The presence of curvature in the electron distribution can remedy this difficulty.

We have applied the equipartition scenario described in a companion paper \citep{equipartitionpaper}, but improved by the inclusion of multiple emission lines in the BLR and use of blackbody rather than monochromatic dust spectrum, to model the SED of the well-studied FSRQ \threec.  Here we have focused our attention on the origin of the break observed in the \fermilat\ spectrum of this object. By introducing curvature in the particle energy distribution, and considering parameters near equipartition, we have shown that the EIC scattering of photons from the dusty torus and BLR can  satisfactorily reproduce the \fermilat\ data in two different flux states. 

In contrast to other EIC models for this source \citep{Bonnoli11, Ghisellini10}, where the feature appears by choosing $\delta$, $B$, and parameters describing the electron spectrum, we show that the GeV break is a natural consequence of BLR radiation scattered by electrons in a blazar jet where near-equipartition conditions hold. In this scenario, the location of the $\gamma$-ray emitting region is constrained to be \textbf{$\gtrsim 10^4$} gravitational radii, where the direct accretion disk radiation makes a weak scattered flux. For \threec, like  \threecbis,  this indicates an  emission region outside the conventional BLR, though still in an environment
with significant emission-line and dust radiation.

\acknowledgments{The work of C.D.D. is supported by the Office of Naval Research and the Fermi Guest Investigator Program.}

\end{document}